\documentclass[aip,graphicx]{revtex4-1}

\usepackage{latexsym,alltt,textcomp,hyperref,color,here,rotating}

\newcommand{\ds}{\displaystyle}

\newcommand{\dd}{{\mathrm d}}

\begin{document}

\title{Dipole-dipole polarizability of the cadmium $\mathbf{^1S_0}$ state revisited}

\author{Uwe Hohm} 
\email{u.hohm@tu-braunschweig.de}

\affiliation{Technische Universit\"at Braunschweig, Institut f\"ur Physikalische und Theoretische Chemie, Gau{\ss}str. 17, D-38106 Braunschweig, Germany}
\email{u.hohm@tu-braunschweig.de}

\begin{abstract}
Experimental results and high-level quantum chemical ab initio calculations of the static polarizability $\alpha_0=\alpha(\omega=0)$ of the cadmium ${^1S_0}$ state are still in marked disagreement. Here we analyze this discrepancy by using experimentally determined dipole oscillator strength distributions (DOSD). It will be shown that within this procedure the experimentally determined static polarizability $\alpha_0$ will shift from $49.7 \pm 1.6~$au to considerably lower values. We now conclude an experimentally determined polarizability of $\alpha_0=47.5 \pm 2.0~$au in much better agreement with the latest calculations of $\alpha_0 \approx 46~$au. 
\end{abstract}


\maketitle

\section{Introduction}
\label{intro}
During the last decades quantum-chemical methods have improved dramatically and in many aspects are now able to match or even surpass the quality of experimental findings of various fundamental atomic and molecular properties \cite{Lach2004,Puchalski2011,Czachorowski2020,Hellmann2021}. This is especially the case for atomic and molecular dipole-dipole polarizabilities $\alpha$ (from now on $\alpha$ is just called the polarizability). For atoms and small molecules experimental and theoretical results nowadays often converge, even in delicate cases such as atomic mercury, see e.g. \cite{Goebel1996hg,Schwerdtfeger2018,Kumar2021}. 
One important exception seems to be the case of atomic cadmium in its ${^1S}_0$ ground state. Cadmium is interesting in view of its possible application for atomic clock standards \cite{Dzuba2019,Guo2021}. However, the latest quantum-chemical calculations of $\alpha$ and the very few experimental results generally do not overlap within their error bars\cite{Sahoo2018,Dzuba2019,Guo2021,Cuthbertson1907,Goebel1995,Goebel1995a}. A marked mismatch of unclear origin remains. Unfortunately we are not able to redo our refractive index measurements from which the polarizability of cadmium was extracted \cite{Goebel1995}. Hence, we are looking for an alternative way to determine the static polarizability $\alpha_0$ without resorting to measured refractivities and measured as well as calculated polarizabilities.
One way of doing this is to deduce the polarizability from experimentally determined dipole oscillator strength distributions (DOSD) \cite{Fano1968,Zeiss1977}. This technique has a long history and it has been shown in numerous examples that it is able to give very reliable data of static and frequency-dependent polarizabilities as well as isotropic and anisotropic dispersion interaction energy coefficients, see e.g. Thakkar \cite{Thakkar2020} and references therein. 
Our idea now is to obtain the static polarizability of cadmium from experimentally determind DOSD data and hopefully to discern between the older experimental and the latest theoretical results of $\alpha_0$ of atomic cadmium. In contrast to the work of Reshetnikov et al \cite{Reshetnikov2008} not only a single transition but the whole spectral distribution of the photoabsorption cross section of atomic cadmium is taken into account.

\section{Dipole properties}
We follow the general treatment of Fano and Cooper \cite{Fano1968} and Berkowitz \cite{Berkowitz1979,Berkowitz2002}, more specific details are given by Kumar and Meath \cite{Kumar1985}. The properties of interest are the dipole sums $S(k)$. They are related to the dipole oscillator strengths (discrete part of the spectrum) $f_i$ and differential dipole oscillator strengths (continuous part of the spectrum) $\dd f / \dd E$ via \cite{Kumar1985}
\begin{equation}
	S(k)=\sum_i \left(\frac{\ds E_i}{\ds 2 R_\infty}\right)^k f_i + \int\limits_{I_p}^\infty \left(\frac{\ds E}{\ds 2 R_\infty} \right)^k \left( \frac{\ds \dd f}{\ds \dd E}\right) \dd E = S_d(k) + S_c(k) \ \\ ,
	\label{DOSD1}
\end{equation} 
$E_i$ is the transition energy, $I_p$ the ionization energy, and $R_\infty$ the Rydberg constant. All energies are given in $\mathrm{m^{-1}}$, The dipole sums $S(k)$ are obtained in atomic units. The first term $S_d(k)$ on the rhs of Eq.(\ref{DOSD1}) is the discrete part and evaluated for energies $E_i < I_p$. The second term $S_c(k)$ is the continuous part of the dipole sum $S(k)$. Due to experimental limitations the discrete part can be summed up only to a certain level $n$ with energy $E_n$ which is very close to the ionization limit $I_p$. In the range $E_n < E \leq I_p$ the quasi-continuous contribution $\Delta S(k)$ to the discrete part of $S(k)$ is approximated via 
\begin{eqnarray}
	S_d(k)&=&\sum_i \left(\frac{\ds E_i}{\ds 2 R_\infty}\right)^k f_i = \sum_i^n \left(\frac{\ds E_i}{\ds 2 R_\infty}\right)^k f_i + \Delta S(k) \label{Sion0} \\ && \nonumber \\
\Delta S(k) &=& \int\limits_{E_\mathrm{n}}^{I_p} \left(\frac{\ds E}{\ds 2 R_\infty} \right)^k \left( \frac{\ds \dd f}{\ds \dd E}\right)\dd E  \approx (2 R_\infty)^{-k}\frac{\ds 1}{\ds k+1}\left(\frac{\ds \dd f}{\ds \dd E}\right)_\mathrm{ion}  \left(I_p^{k+1} - E_\mathrm{n}^{k+1}\right)
	\label{Sion1}
\end{eqnarray}
$(\dd f/ \dd E)_\mathrm{ion}$ can be obtained from photoionization cross sections\cite{Ahmad1996}. According to the Kuhn-Reiche-Thomas (KRT) sum rule for a $N-$electron system $S(0)=N$ \cite{Kuhn1925,Reiche1925,Fano1968}. The static polarizability is related to $S(k)$ via $\alpha_0=S(-2)$. Therefore $\alpha_0$ can be obtained by evaluating the dipole sums given in Eq. \ref{DOSD1}.

\section{Input Data}
\label{sec:1}
\subsection{Discrete photoabsorption cross section} \label{Input-Ds}
Input data for the discrete part of the spectrum are given in Table \ref{Input-D}. If necessary, transition probabilities $A$ are converted to oscillator strengths $f$ via $g_i f=1.499 \times 10^{-8} \lambda^2 g_k A$, where $\lambda$ is given in {\AA}ngstr\"oms, $A$ in $10^8/~$s, and $g_i$ and $g_k$ are the degeneracies of the corresponding electronic states\cite{CRC2010}. Due to lack of experimental data the oscillator strengths for the two transitions $\mathrm{5 s^2~ {^1S}_0 \to 5s~np~ {^1P}_1}$ with $n=6$ and $n=7$ are estimated from calculated results of $K_n=f_n (n^*)^3/(2 R_\infty)$ which are presented in Fig. 3 of Ref.\cite{Ahmad1996}. For $n$=5, 6, and 7 we extract $K_n$ as $2.6 \times 10^{-3}$, $0.88 \times 10^{-3}$, and $0.63 \times 10^{-3}$, respectively (all in $\mathrm{m^{-1}}$). Setting $n^*=n-3$ we obtain $f_5=0.713$, $f_6=0.0715$, and $f_7=0.0216$. However, we do not use these data directly. Instead of this we fix $f_5$ to the experimental value of $f_5=1.3442$ \cite{Xu2004}. This experimental value is larger by a factor of 1.89 than the calculated one of 0.713. Therefore, we have scaled the calculated data of $f_6$ and $f_7$ by the same factor and obtain $f_6=0.135$ and $f_7=0.041$, see Table \ref{Input-D}. 

\begin{table}
\caption{\label{Input-D} $\mathrm{5 s^2~ {^1S}_0 \to 5s~5p~ {^3P}_1}$ (first row) and $\mathrm{5 s^2~ {^1S}_0 \to 5s~np~ {^1P}_1}$ discrete oscillator strengths $f_i$ and transition frequencies $E_i$ (given in $\mathrm{m^{-1}}$) of cadmium.}
\begin{tabular}{cccc} \\ \hline
$n$&$E_i$ / $\mathrm{m^{-1}}$&$f_i$& Reference\\ \hline
5'&3065609&0.0019&Ref. \cite{Ashenfelter1967}\\
5&4370629&1.3442&Ref.\cite{Xu2004}\\
6&5990728&0.135&$E_i$ Ref.\cite{Brown1975}, $f_i$ see text\\
7&6550141&0.041&$E_i$ Ref.\cite{Brown1975}, $f_i$ see text \\
8&6805940&0.0143&Ref. \cite{Ahmad1996} \\
9&6943910&0.0069&Ref. \cite{Ahmad1996} \\
10&7026740&0.0038&Ref. \cite{Ahmad1996} \\
11&7080330&0.0024&Ref. \cite{Ahmad1996} \\
12&7116970&0.0016&Ref.\cite{Ahmad1996} \\
13&7143140&0.0012&Ref. \cite{Ahmad1996} \\ \hline
\end{tabular}
\end{table}

\subsection{Continuous photoabsorption cross section} \label{Input-C}
Input data in the range between $10.2 \leq E/\mathrm{eV} \leq 29779$~eV are taken from Henke et al. \cite{Henke1993}, in the range $29779 < E/\mathrm{eV} \leq 433000$ photoabsorption cross sections compiled by Chantler \cite{Chantler1995} are used. The photoionization cross section of $\sigma=4 \pm 2~$Mbarn at $I_p=8.9938~$eV is taken from Ahmad et al. \cite{Ahmad1996}.
 
\begin{figure}[h]
\includegraphics[width=1.0\linewidth]{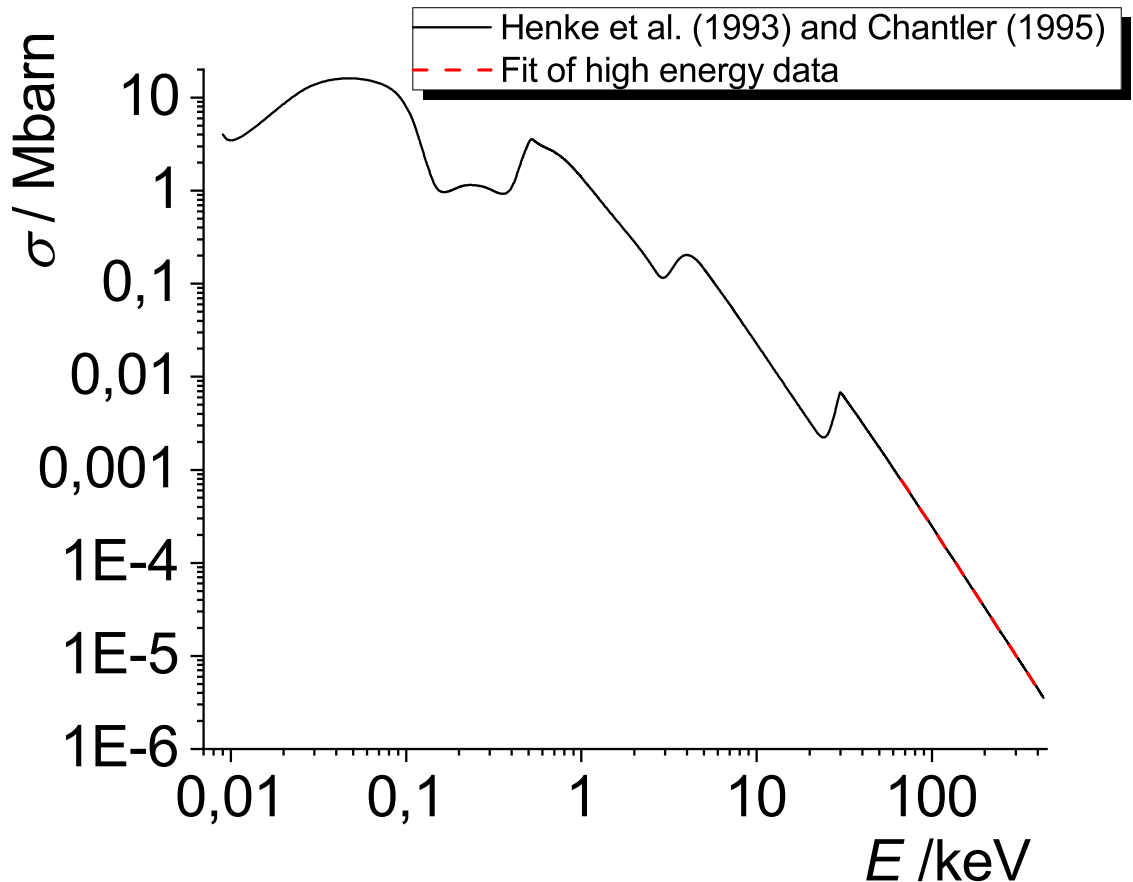}%
\caption{\label{Continuum-pic}Graphical representation of high-energy cadmium photoabsorption cross-sections (black curve)\cite{Henke1993,Chantler1995}. At high energies, the fit (dashed red line) coincides nearly exactly with the recommended data given by Chantler\cite{Chantler1995}.} 
\end{figure}

\begin{table}
\caption{\label{Output-all} Contributions from the discrete and continuous spectrum of cadmium to $S(k)$. The discrete transitions are listed in Table \ref{Input-D}}
\begin{tabular}{cccc} \\ \hline
Transition&$S(0)$&$S(-2)$&Remarks \\ \hline
5'&0.0019&0.0974&\\
5&1.3442&33.8957&\\
6&0.135&1.8119&\\
7&0.041&0.4603& \\
8&0.0143&0.1487& \\
9&0.0069&0.0689& \\
10&0.0038&0.0371\\
11&0.0024&0.0231& \\
12&0.0016&0.0152& \\
13&0.0012&0.0113 \\
$\Delta S(k)$&0.0050&0.0466&See Eq.(\ref{Sion1})\\ \hline
$S_d(k)$&1.5573&36.6162& \\ 
$S_c(k)$&43.3976&5.6657&\\ 
$S(k)=S_d(k)+S_c(k)$&44.9549&42.2819& This work, raw data \\
$S(k)=x\times S_d(k)+y\times S_c(k)$&47.9829&47.5201&This work, scaled, see text \\
Theoretical&48~(KRT)&46.02(50), 46.52, 45.92(10)  &Refs.\cite{Sahoo2018,Dzuba2019,Guo2021} \\
Experiment&&$\approx 50$, 49.65(1.62), 45.31(1.7)&Refs.\cite{Cuthbertson1907,Goebel1995,Goebel1995a} \\  \hline
\end{tabular}
\end{table}

\begin{figure}
	\includegraphics[width=1.0\linewidth]{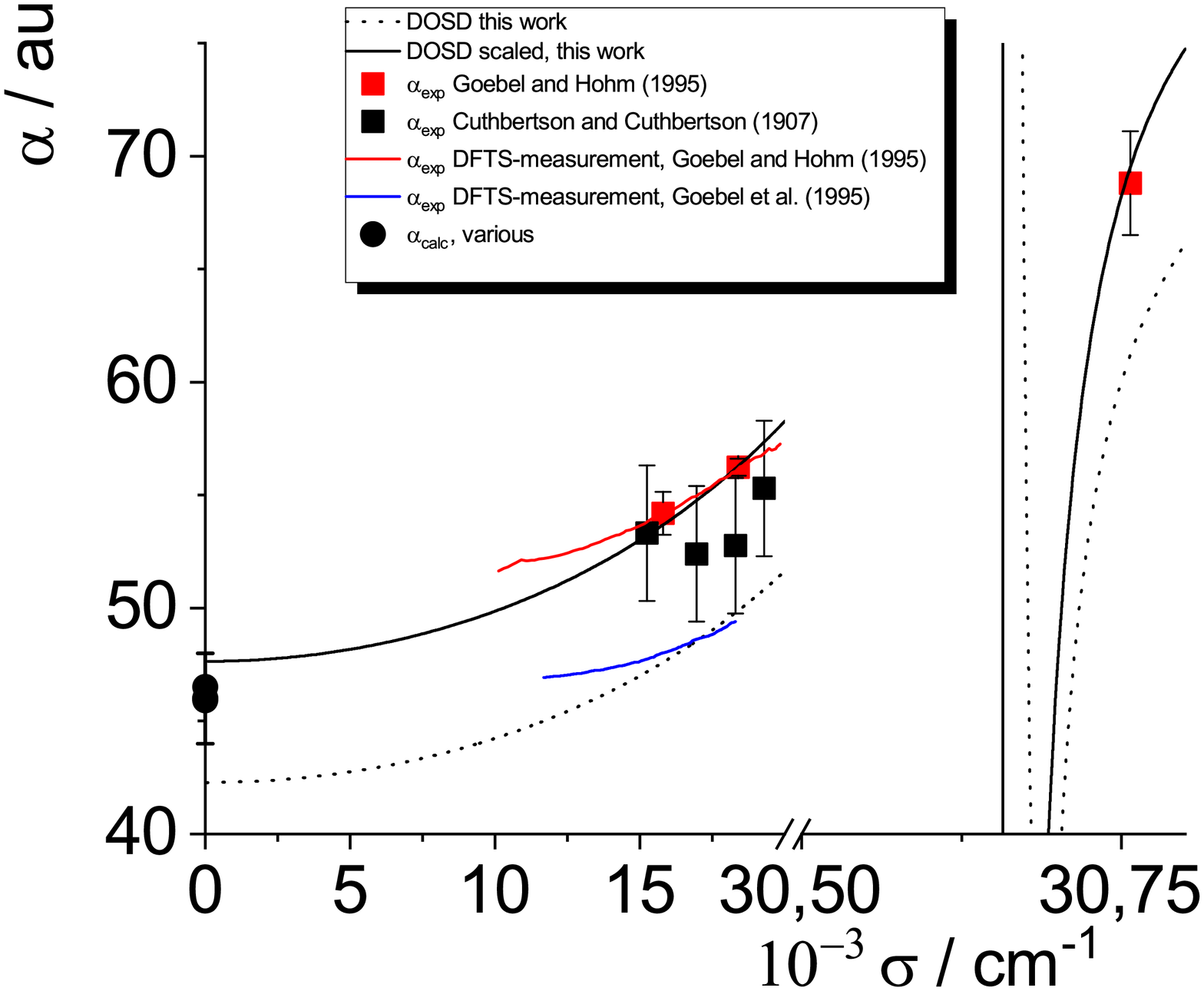}%
	\caption{\label{Pol-all}Comparison of polarizabilities $\alpha(\lambda)$ of atomic cadmium obtained from DOSD (black dotted line), scaled DOSD (black full line), monochromatic experiments (red squares Ref.\cite{Goebel1995}, black squares Ref.\cite{Cuthbertson1907}), DFTS-experiments (red line\cite{Goebel1995}, blue line\cite{Goebel1995a}), and quantum chemical calculations (black circles, Refs.\cite{Schwerdtfeger2018,Sahoo2018,Dzuba2019,Guo2021}). Note that DFTS-measurements are relative measurements which are fitted to a single-wavelength measurement of the polarizability. } 
\end{figure}

\section{Results and Discussion}
The discrete part $S_d(k)$ of the dipole sums $S(k)$ is calculated via Eqs. (\ref{Sion0}) and (\ref{Sion1}). $( \dd f/ \dd E)_\mathrm{ion}=4.5198 \times 10^{-8}~$m is used \cite{Ahmad1996}. The resulting values of the dipole sums $S(0)$ and $S(-2)$ are given in Table \ref{Output-all}. 
Integration of the continuous part in Eq. (\ref{DOSD1}) is carried out via a non-equidistant form of the Simpson formula \cite{Brun1953} in the range between $I_p=7254007~\mathrm{m^{-1}}$ and $E_\mathrm{max}=8.60069 \times 10^{10}~\mathrm{m^{-1}}$ ($\approx 107$~keV). For $E>E_\mathrm{max}$ $(\dd f / \dd E)$ is fitted to $(\dd f / \dd E)=A\times (E / \mathrm{m^{-1}})^B$. We obtain $A=8.7689\times 10^{19}~$m and $B=-2.8884$, see Fig.\ref{Continuum-pic}. Subsequently, the integration for $E\geq E_\mathrm{max}$ is carried out analytically. However, its contribution to $S(k)$  in the case of $k=0$ and $k= - 2$ is found to be practically negligible. The contribution of the continuous part of the photoabsorption spectrum is also given in Table \ref{Output-all}.
Taking into account both contributions we obtain for the sum $S(k)=S_d(k)+S_c(k)$ in total $S(0)=44.95$ and $S(-2)=42.28~$au. Both values are near the expected results of $S(0)=N=48$ and $S(-2)=(46 \ldots 50)~$au. However both are smaller indicating that the experimental input data are either incomplete and/or slightly wrong. This behaviour is nearly expected for heavy atoms where only in very rare cases sufficient data of high quality are available in order to obtain reliable values of the $S(k)$ \cite{Berkowitz1979,Reshetnikov2008}. At this point we cannot conclude whether the experimentally determined static polarizability of $\alpha_0 \approx 49.7~$au or the calculated value of $\alpha_0 \approx 46.0~$au is more reliable. 
We proceed by calculating the dynamic polarizability $\alpha(\lambda)$ from the experimentally determined photoabsorption data. To this end the terms $E_i^k$ and $E^k$ in Eq.(\ref{DOSD1}) must be replaced by $1/(E_i^2-(1/\lambda)^2)$ and $1/(E^2-(1/\lambda)^2)$, respectively, and the dynamic (wavelength) dependent polarizability is calculated via\cite{Kumar1985}
\begin{equation}
	\alpha(\lambda)=(2 R_\infty)^2 \left(\sum_i \frac{\ds f_i}{\ds E_i^2-(1/\lambda)^2} + \int\limits_{I_p}^\infty \frac{\ds (\dd f / \dd E)}{\ds E^2-(1/\lambda)^2} \dd E \right) \ \\ ,
	\label{DOSD2}
\end{equation}
 $\lambda$ is the wavelength used in the experiments. The resulting curve is displayed in Figure \ref{Pol-all}. It can be seen that the polarizability obtained in this work underestimates the measured and calculated data considerably. As in the case of the static polarizability this is due to insufficient DOSD data\cite{Reshetnikov2008}. This result is disappointing but somewhat expected. 
To carry on with our analysis we have now scaled the two individual contributions $S_d(k)$ and $S_c(k)$ in order to fullfil the KRT sum-rule $S(0)=48$ and to get as close as possible to the measured values of $\alpha(\lambda)$ ($\alpha(632.99~\mathrm{nm})=54.20~$au, $\alpha(543.51~\mathrm{nm})=56.23~$au, and $\alpha(325.13~\mathrm{nm})=68.8~$au\cite{Goebel1995}). If we set $S(k)=x\times S_d(k)+ y \times S_c(k)$ the optimal result is $x=1.133$ and $y=1.065$. The scaled values are also displayed in Figure \ref{Pol-all}. We obtain $S(0)=47.98 \approx N=48$ and the static (zero frequency) polarizability as  $\alpha_0=S(-2)=47.52$. $\alpha_0$ now is much closer to the theoretical results, see Table \ref{Pol-all}. Of course, $\alpha(\lambda)$ now fit our own experimental data \cite{Goebel1995} very well. But the scaled curve is also inside the error bars of the meaurements of Cuthbertson and Cuthbertson \cite{Cuthbertson1907}. 
In the last step of our analysis we compare our findings with polarizability data obtained from dispersive Fourier transform spectroscopy (DFTS), which is a quasi-continuous white light interferometric technique in the visible wavelength range. With this special technique the polarizability is not recorded at a single wavelength but in a broad wavelength interval. However, results obtained with DFTS are only relative to a given absolute value $\alpha(\lambda_1)$ recorded at a wavelength $\lambda_1$. This technique has been used very successfully for measuring the quasi-continuous polarizability spectrum of $\mathrm{NO_2/N_2O_4}$, $\mathrm{C_{10}H_{16}}$ (adamantane), $\mathrm{Fe(C_5H_5)_2}$ , $\mathrm{Ru(C_5H_5)_2}$, $\mathrm{Os(C_5H_5)_2}$, $\mathrm{Nd(C_5H_5)_3}$ , $\mathrm{Sm(C_5H_5)_3}$, $\mathrm{Er(C_5H_5)_3}$, $\mathrm{P_4}$, $\mathrm{As_4}$, Zn, and Hg \cite{Goebel1994,Maroulis2001,Goebel1997farad,Hohm2001cpl,Hohm2000p4,Hohm2000rsi,Goebel1996as4,Hohm1998as4,Goebel1996zn,Goebel1996hg}. Even in the case of strongly absorbing iodine vapour the experimentally measured polarizability spectrum inside the absorption band is in very good accordance with theoretical calculations \cite{Maroulis1997,Hohm1999i2,Hohm2000rsi}. In summary all of the hitherto performed white-light interferometric measurements (DFTS) fit extremely well to other experimental and theoretical findings. However, in the case of cadmium the situation seems to be totally different. The white light interferometric (DFTS) measurements of Goebel and Hohm \cite{Goebel1995} and Goebel et al. \cite{Goebel1995a} do not fit to the findings obtained in this work from DOSD, see Figure \ref{Pol-all}. The red curve \cite{Goebel1995} is a mean of 13 measurements, where the blue curve\cite{Goebel1995a} is a single measurement only. The latter one was our first recording of the polarizability spectrum of a metal vapor. Therefore it should be regarded of much lower quality than $\alpha(\lambda)$ presented in the red curve. It is very important to note that in the visible wavelength range the slope of the DFTS measurements is much smaller than that obtained from our DOSD or scaled DOSD analysis. Therefore the DFTS curves match our single wavelength measurements at 632.99~nm and 543.51~nm but do not fit to our resulting (scaled) curves obtained from DOSD. Atomic cadmium seems to be the very first case where a severe mismatch between the experimentally determined frequency dependence of $\alpha(\lambda)$ and other spectroscopic properties is observed. It is very likely that our former extrapolation of the DFTS data (red curve in Figure \ref{Pol-all}) to zero frequency result in an erroneous value of the static polarizability of the cadmium ${^1S_0}$-state. We suggest a new value which is in accord with our single frequency measurements and the DOSD analysis presented in this paper of $\alpha_0=\alpha(\omega=0)=47.5 \pm 2.0~$au, bearing in mind that this extrapolation disagrees with DFTS measurements. It must be stressed that the origin of this observed discrepancy remains unclear.

\section{Conclusion}
We try to use experimentally determined dipole oscillator strength distributions for unravelling the discrepancy between the experimental and theoretical static dipole-dipole polarizability $\alpha_0$ of the atomic cadmium ${^1S_0}$-state. However, the photoabsorption data given in the literature obviously cannot totally account for the theoretically predicted dipole sums $S(0)=N$ and $\alpha_0=S(-2)$. Therefore, by using only these data sets it is not possible to discern between the experimental and calculated $\alpha_0$ of cadmium. Nevertheless a comparison with polarizability spectra obtained from DFTS measurements indicates that some parts of our former measurements\cite{Goebel1995,Goebel1995a} might be in error. However, the reason for this is unclear. Taking the present results into account we suggest a new experimental value of the static polarizability of Cd of $\alpha_0=47.5 \pm 2.0~$au. $\alpha_0$ now is in much better agrement with the theoretical result of $\alpha_0 \approx 46~$au. This new experimental value might be checked by using totally different experimental techniques like time-of-flight investigations of laser cooled Cd atoms in electric fields\cite{Amini2003}.


\bibliography{Hohm-bib-Cd-v0}   

\end{document}